\documentclass[a4paper,11pt]{article}
\usepackage{pos}

\usepackage{adjustbox}

\newcommand{\be}{\begin{equation}}
\newcommand{\ee}{\end{equation}}
\newcommand{\bea}{\begin{eqnarray}}
\newcommand{\eea}{\end{eqnarray}}
\newcommand{\nn}{\nonumber}

\newcommand{\np}{\textsc{NP}}

\title{Constraints on new physics couplings from ${\bar B} \to D^*\,(D\,\pi)\,\ell\,{\bar \nu}_{\ell}$ angular analysis}
\author*[a,b]{Nicola Losacco}
\affiliation[a]{Dipartimento Interateneo di Fisica "Michelangelo Merlin", Universit\`a degli Studi di Bari,\\
via Orabona 4, 70126 Bari, Italy}

\affiliation[b]{Istituto Nazionale di Fisica Nucleare, Sezione di Bari,\\
via Orabona 4, 70126 Bari, Italy}

\emailAdd{nicola.losacco@ba.infn.it}
\abstract{The Belle Collaboration has measured the complete set of angular coefficient functions for the decays ${\bar B} \to D^*\,(D\,\pi)\,\ell\,{\bar \nu}_{\ell}$, where $\ell = e,\,\mu$, in four bins of the variable $w={m_B^2+m_{D^*}^2-q^2 \over 2\, m_B\, m_{D^*}}$, with $q$ the momentum of the lepton pair. In SM this measurement is instrumental in determining the hadronic form factors in the $B \to D^*$ matrix elements of the SM weak current, thereby refining the measurement of $\lvert V_{cb} \lvert$. On the other hand, it can be used to assess the impact of possible new physics contributions. I consider an extension of the SM effective Hamiltonian that governs this mode, incorporating the complete set of Lorentz invariant $d = 6$ operators compatible with the gauge symmetry of the theory. The measured angular coefficient functions play a pivotal role in constraining the couplings in the generalized effective Hamiltonian.}

\FullConference{42nd International Conference on High Energy Physics (ICHEP2024)\\
18-24 July 2024\\
Prague, Czech Republic\\}

\begin{document}

\renewcommand{\hookAfterAbstract}{%
\par\bigskip
\textsc{BARI-TH/764-24}
%:\href{https://arxiv.org/abs/1234.5678}{1234.5678}
}
\maketitle

\section{Introduction and framework}
In the search for physics beyond the SM, tensions have emerged in the flavour sector \cite{DeFazio:2023lmy, Colangelo:2024ped}. In  particular, anomalies appeared in processes sensitive to virtual contributions of heavy particles. Anomalies in charged current processes appeared as well, as in the ratios $R(D^{(*)}) = {\mathcal{B}(B \to D^{(*)}\, \tau \,\nu_\tau) \over \mathcal{B}(B \to D^{(*)}\, \ell \, \nu_{\ell})}$ ($\ell = e$ or $\mu$) \cite{HFLAV:2022pwe}. % and have since been investigated in multiple studies \cite{BaBar:2013mob,Belle:2015qfa,LHCb:2015gmp,Belle:2016dyj,LHCb:2017smo,LHCb:2017rln,Belle:2019rba}. 
The possible connection between these results and the discrepancies in various determinations of $|V_{cb}|$ adds further intrigue to the investigation of these processes \cite{Colangelo:2016ymy}. 

The fully differential decay distribution of $\bar B(p) \to D^{(*)}(p^\prime,\,\epsilon)(D \pi) \ell (k_1){\bar \nu}_\ell(k_2)$ is sensitive to the effects of new physics (NP), as well as to hadronic quantities \cite{Colangelo:2018cnj,Bhattacharya:2018kig,Murgui:2019czp,Becirevic:2019tpx,Bobeth:2021lya,Martinelli:2021onb}.Currently, it is under experimental investigation \cite{Belle:2023xgj}. A model-independent framework for studying NP contributions is provided by the Standard Model Effective Field Theory (SMEFT) \cite{Buchmuller:1985jz, Grzadkowski:2010es} . Denoting by $\Lambda_{NP}$ the NP scale and assuming that it is much larger that the weak scale, possible new massive particles/mediators can be integrated out. An effective Hamiltonian at the electroweak (EW) scale is obtained, that extends the SM one with the inclusion of new operators expressed in terms of SM fields and weighted by inverse powers of $\Lambda_{NP}$. At order $1/\Lambda_{NP}^2$ these are $d=6$ operators. Constraints on their coefficients help identifying the features of NP. For example, the possibility to exploit gauge anomaly cancellation in a $U(1)$ extension of the SM has been exploited in \cite{Colangelo:2024sbf}.

The generalized effective Hamiltonian that describes the modes  ${\bar B} \to V \ell^-  {\bar \nu}_\ell$, with $V$ a  meson comprising an up-type quark $U$ has the form
 \bea
H_{\rm eff}^{b \to U \ell \nu} &=& {G_F \over \sqrt{2}} V_{Ub} \times \Big\{(1+\epsilon_V^\ell) \left({\bar U} \gamma_\mu (1-\gamma_5) b \right)\left( {\bar \ell} \gamma^\mu (1-\gamma_5) {\nu}_\ell \right)
\nn \\
&+&\epsilon_R^\ell \left({\bar U} \gamma_\mu (1+\gamma_5) b \right)\left( {\bar \ell} \gamma^\mu (1-\gamma_5) {\nu}_\ell \right)
+  \epsilon_S^\ell \, \left({\bar U}  b\right)  \left({\bar \ell} (1-\gamma_5) { \nu}_\ell \right)  \label{heff} \\
&+& \epsilon_P^\ell \, \left({\bar U} \gamma_5 b\right)  \left({\bar \ell} (1-\gamma_5) { \nu}_\ell \right) + \epsilon_T^\ell \, \left({\bar U} \sigma_{\mu \nu} (1-\gamma_5) b\right) \,\left( {\bar \ell} \sigma^{\mu \nu} (1-\gamma_5) { \nu}_\ell \right) \Big\} + h.c.\,\,\, , \nn
\eea
with $G_F$ the Fermi constant and $V_{Ub}$ the Cabibbo-Kobayashi-Maskawa (CKM) matrix element involved in the process. In  \eqref{heff}, the SM term as well as NP operators with complex  $\epsilon^\ell_{V,R,S,P,T}$  lepton-flavour dependent coefficients appear. The scalar operator does not contribute if $V$ is a vector meson, as $D^*$ considered in the present case.
\begin{figure}[!h]
\begin{center}
\includegraphics[width = 0.4\textwidth]{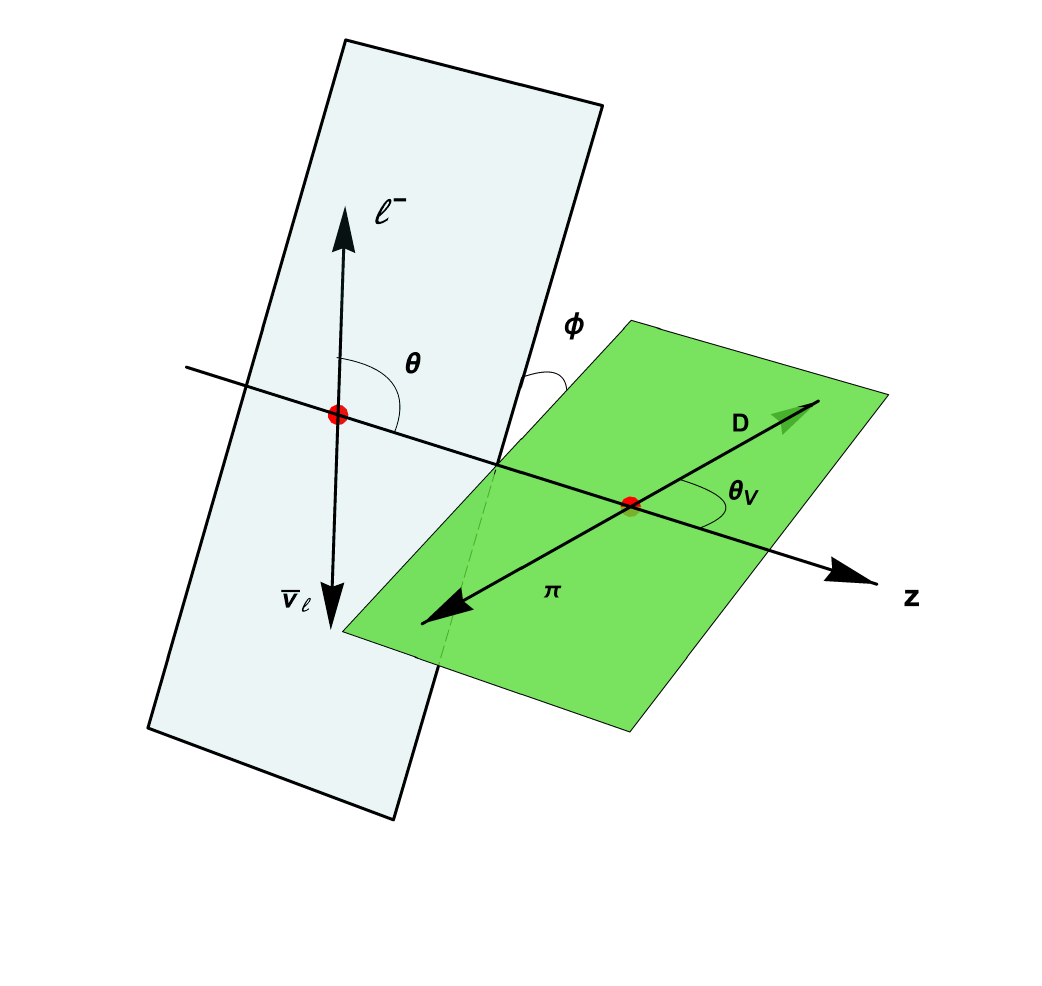}\vspace*{-1.cm}
\caption[Kinematics of  $\bar B \to D^*(D\pi) \ell^- \bar \nu_\ell$.]{  \small  Kinematics of  $\bar B \to D^*(D\pi) \ell^- \bar \nu_\ell$.% $\theta_V$ is the angle between the $D$ meson and the direction opposite to the ${\bar B}$ in the $D^*$ rest frame; $\theta_\ell$ the angle between the charged lepton and the ${\bar B}$ in the virtual $W$ rest frame and $\phi$ the angle between the decay planes identified by the directions of the lepton pair on one hand and the $(D,\pi)$ pair on the other in the ${\bar B}$ rest frame.
}\label{Fig:angles}
\end{center}
\end{figure}

Choosing the kinematics as in Fig.~\ref{Fig:angles} the fully differential decay width reads:
\bea
\frac{d^4 \Gamma (\bar B \to V( P_1 P_2) \ell^- \bar \nu_\ell)}{dq^2 \,d\cos \theta \,d\phi \,d \cos \theta_V} 
&=&{\cal C}|{\vec p}_{V}| \left(1-  \frac{ m_\ell^2}{q^2}\right)^2 \times \Big\{I_{1s} \,\sin^2 \theta_V+I_{1c} \,\cos^2\theta_V \nn \\
&+&\left(I_{2s} \,\sin^2 \theta_V+I_{2c} \,\cos^2 \theta_V\right) \cos 2\theta
+ I_3 \,\sin^2 \theta_V \sin^2 \theta  \cos 2 \phi \nn \\
&+& I_4 \, \sin 2\theta_V \sin 2\theta \cos  \phi + I_5 \, \sin 2 \theta_V \sin \theta \cos  \phi \label{angular} \\
&+&\left(I_{6s} \,\sin^2 \theta_V+I_{6c} \,\cos^2\theta_V\right)\cos \theta + I_7 \sin 2 \theta_V \sin \theta \sin  \phi \nn \\
&+& I_8 \,\sin 2 \theta_V \, \sin 2\theta \sin  \phi + I_9\,\sin^2 \theta_V \sin^2 \theta \sin  2\phi
  \Big\}\,\,, \nn
\eea
with ${\cal C}=\displaystyle{\frac{3G_F^2 |V_{Ub}|^2 {\cal B}(V \to P_1 P_2)}{128(2\pi)^4m_B^2}}$ and ${\vec p}_{V}$  the three-momentum of the $V$ meson ($D^*$) in the $B$ rest-frame.
The angular coefficient functions $I_i$ in \eqref{angular} depend on the couplings $\epsilon^\ell_{V,R,P,T}$, on $q^2$ (or $w$) and on the hadronic form factors in the $B \to V$ matrix elements. They read
\bea
I_i &=& |1+\epsilon_V|^2 \,I_i^{SM}+|\epsilon_R|^2I_i^{NP,R}+|\epsilon_P|^2I_i^{NP,P} + |\epsilon_T|^2I_i^{NP,T} +2 \, {\rm Re}\left[\epsilon_R(1+\epsilon_V^* )\right] I_i^{INT,R} \nn \\
&+&2 \, {\rm Re}\left[\epsilon_P(1+\epsilon_V^* )\right] I_i^{INT,P} + 2 \, {\rm Re}\left[\epsilon_T(1+\epsilon_V^* )\right] I_i^{INT,T} + 2 \, {\rm Re}\left[\epsilon_R \epsilon_T^* \right] I_i^{INT,RT}  \label{eq:Iang1} \\
&+& 2 \, {\rm Re}\left[\epsilon_P \epsilon_T^* \right] I_i^{INT,PT} + 2 \, {\rm Re}\left[\epsilon_P \epsilon_R^* \right] I_i^{INT,PR} \nn
\eea
for $ i=1,\dots 6$,

\begin{equation}
\begin{aligned}
I_7 &=2 \, {\rm Im}\left[\epsilon_R(1+\epsilon_V^* )\right] I_7^{INT,R} + 2 \, {\rm Im}\left[\epsilon_P(1+\epsilon_V^* )\right] I_7^{INT,P} + 2 \, {\rm Im}\left[\epsilon_T(1+\epsilon_V^* )\right] I_7^{INT,T}  \\
&+ 2 \,{\rm Im}\left[\epsilon_R \epsilon_T^* \right] I_7^{INT,RT}+2 \, {\rm Im}\left[\epsilon_P \epsilon_T^* \right] I_7^{INT,PT} + 2 \, {\rm Im}\left[\epsilon_P \epsilon_R^* \right] I_7^{INT,PR}  , \label{eq:Iang2} 
\end{aligned}
\end{equation}
and for $ i=8,\,9$
\be
I_i=2 \, {\rm Im}\left[\epsilon_R (1+\epsilon_V^* ) \right] I_i^{INT,R} \,\,\, . 
\label{eq:Iang3} \ee
Their expressions can be found in \cite{Colangelo:2024mxe}.

%In SM such functions are expressed in terms of helicity amplitudes:
%%
%\bea
%H_0 &=&\frac{1}{{2m_V(m_B+m_V) \sqrt{q^2}}}\,\Big( (m_B+m_V)^2(m_B^2-m_V^2-q^2) A_1(q^2) - \lambda(m_B^2,\,m_V^2,\,q^2) A_2(q^2) \Big) \nn  \\
%H_\pm&=& \frac{(m_B+m_V)^2 A_1(q^2)\mp\sqrt{\lambda(m_B^2,\,m_V^2,\,q^2)}V(q^2)}{m_B+m_V}  \label{HampV} \\
%H_t&=& -\frac{\sqrt{\lambda(m_B^2,\,m_V^2,\,q^2)}}{\sqrt{q^2}} \,A_0(q^2) , \,\,\,  \nn
%\eea
%with the form factors defined in the appendix A in \cite{Colangelo:2024mxe}.
%For NP operators the amplitudes are also introduced:
%\begin{equation}
%\begin{aligned}
%&H_\pm^{NP} =
% \frac{1}{\sqrt{q^2}}\Big\{q^2(T_1(q^2)- T_2(q^2)) + \Big(m_B^2-m_V^2 \pm \sqrt{\lambda(m_B^2,m_V^2,q^2)} \Big)(T_1(q^2)+ T_2(q^2))\Big\} \\
%&H_L^{NP}= 4\Big\{
%\frac{\lambda (m_B^2,m_V^2,q^2)}{m_V(m_B+m_V)^2} \, T_0(q^2) + 2\frac{m_B^2+m_V^2-q^2}{m_V} T_1(q^2)+4 m_V T_2(q^2)\Big\}  . \label{HampNP}
%\end{aligned}
%\end{equation}
%The  form factors $T_i$ are also defined in appendix A in \cite{Colangelo:2024mxe}. The  expressions of all coefficient functions $I_i$ are  in the appendix B in \cite{Colangelo:2024mxe}.

\section{Constraints on NP couplings from  Belle measurement}
\begin{figure}[!h]
\begin{center}
\includegraphics[width = 0.35\textwidth]{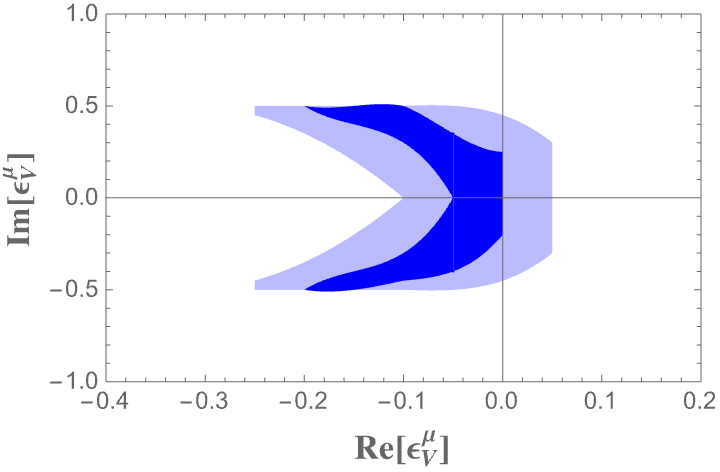}
\includegraphics[width = 0.35\textwidth]{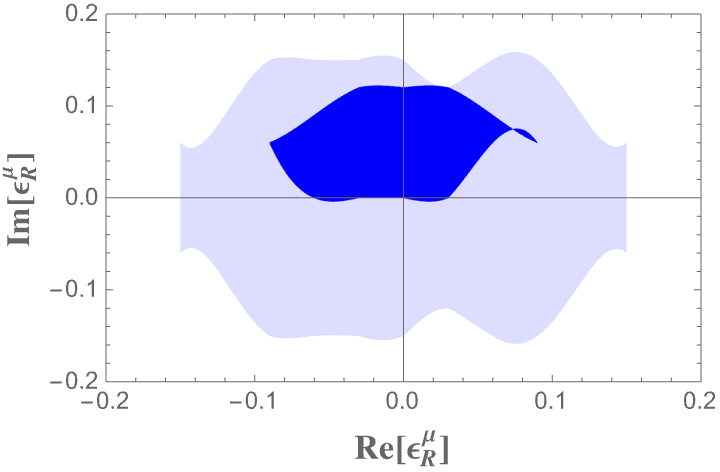}\\
\includegraphics[width = 0.35\textwidth]{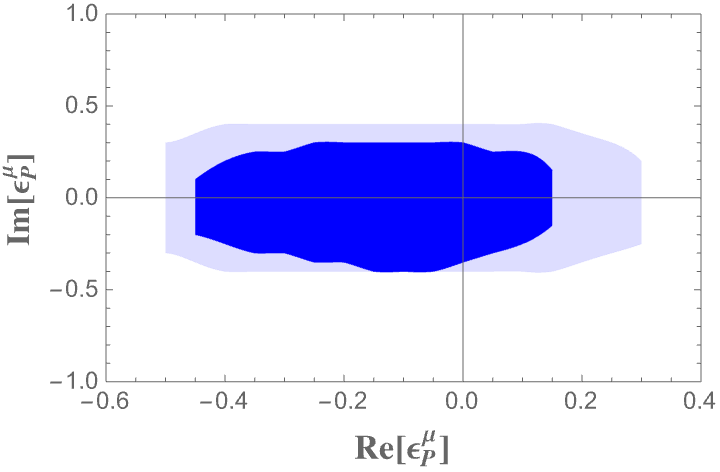}
\includegraphics[width = 0.35\textwidth]{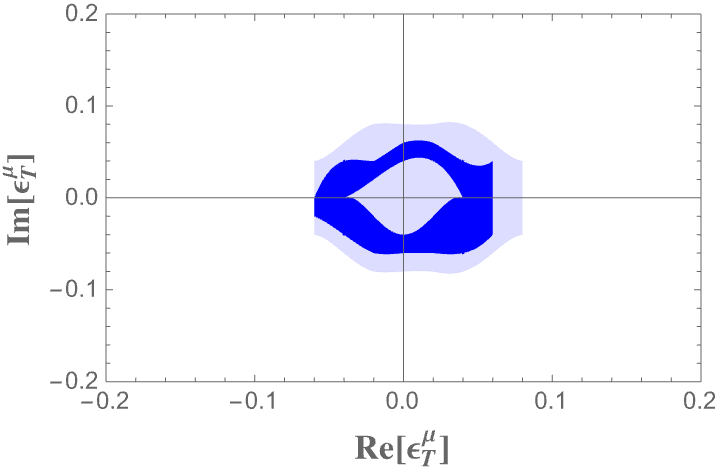}
\caption[ NP couplings parameter space.]{\small NP couplings  $\epsilon_V^\mu,\, \epsilon_R^\mu,\,\epsilon_P^\mu,\,\epsilon_T^\mu$ obtained from the Belle measurement of  the angular coefficient functions of  $\bar B \to D^*(D\pi) \mu^- \bar \nu_\mu$. The light region corresponds to requiring that the theoretical results agree with the experimental data at  2.5 $\sigma$. The dark region corresponds to the values obtained  minimizing the reduced $\chi^2$. }\label{Fig:epsilon}
\end{center}
\end{figure}
Using Belle measurement of the angular coefficient functions \cite{Belle:2023xgj}, it is possible to constrain the couplings in \eqref{heff}. The correspondence of the measured quantities and the used observables is discussed in \cite{Colangelo:2024mxe}. The Belle Collaboration presents angular coefficient functions in four bins $\Delta w^{(a)}$ of the hadronic recoil parameter $w = {m_B^2 + m_{D^{*}}^2 - q^2 \over 2\,m_B \, m_{D^*}}$, with $q = p- p^\prime$. The binned coefficients are defined as ${\bar J}_i^a=\int_{{\Delta w}^{(a)}} J_i(w) dw$, and normalized as  ${\hat J}_i=J_i/{\cal N}$, where ${\cal  N}$ is a normalization factor corresponding, up to a constant, to the integrated width. The values of the coefficients ${\hat J}_i$ in the four bins ${\Delta w}^{(1)}=[1,\, 1.15]$, ${\Delta w}^{(2)}=[1.15,\, 1.25]$, ${\Delta w}^{(3)}=[1.25,\, 1.35]$ and ${\Delta w}^{(4)}=[1.35,\, 1.5]$ can be extracted from Fig.~1 of \cite{Belle:2023xgj}. The products $({\hat J}_i^a )^{\rm exp}_{\rm int}={\hat J}_i \cdot (\Delta w)^a$ are computed. Using the expressions in \cite{Colangelo:2024mxe} and setting ${\cal N}=0.146$ to reproduce the measured branching ratio \cite{Navas:PDG}, we compute the integrals of the coefficient functions for each bin $w$, $({\hat J}_i^a)^{\rm th}_{\rm int}$. We impose the condition $({\hat J}_i^a)^{\rm th}_{\rm int} \in [({\hat J}_i^a )^{\rm exp}_{\rm int} - k\, \sigma_i^a,\, ({\hat J}_i^a )^{\rm exp}_{\rm int} +k\, \sigma_i^a]$, where $k$ is the number of standard deviations and $\sigma_i^a$ is the error of the Belle result for each ${\hat J}_i^a$ multiplied by the bin width. This provides 48 constraints, one for each of the four bins of the 12 angular coefficients. Focusing on the muon channel, we determine the set of parameters $(\epsilon_V^\mu,\, \epsilon_R^\mu,\,\epsilon_P^\mu,\,\epsilon_T^\mu)$ that satisfies all constraints within the initial ranges $|\epsilon_i^\mu|\le 0.5$ for $i=V,\,R,\,P\, ,T$ (set 1). The smallest value of $k$ that satisfies all constraints is $k=2.5$.

The reduced chi-square $\chi^2_{red}=\displaystyle\frac{1}{N_{dof}}\sum_{i,\,a}{ \left( ({\hat J}_i^a)^{\rm th}_{\rm int}-({\hat J}_i^a)^{\rm exp}_{\rm int} \right) ^2}/{\left (\sigma_i^a \right)^2}$, with $N_{dof}=40$,  $i=1s,...9$ and $a=1,...4$, lies between $1.8$ and $2.6$. We select points where $\chi^2_{red}\le  1.875$ (set 2). The probability of finding $\chi^2_{red}>1.875$ with 40 degrees of freedom is $0.07\%$. For the SM (where 
 $\epsilon_V=\epsilon_R=\epsilon_P=\epsilon_T=0$ and $N_{dof}=48$) we find $\chi^2_{red}=2$.

The results are in Fig.~\ref{Fig:epsilon}:  the light region corresponds to the parameters in set 1, the dark region to the parameters in set 2.
% The red point corresponds to the smallest  value of $\chi^2$ among the parameters in set 2.}
%
A fit of the Belle points with the angular coefficient functions obtained using the determined $\epsilon$ couplings is in Fig. 3 of \cite{Colangelo:2024mxe} .
%In Fig.~\ref{angcoeff} the Belle points together with the angular coefficient functions obtained using the determined $\epsilon$ couplings are shown. 
The remarkable result is that, while the SM point with all new couplings equal to zero is allowed, it does not belong to the region of minimum $\chi^2$, there is the possibility of non-vanishing $\epsilon_T^\mu$.
%
%This observation will be strengthened when the table of measurements and the error covariance matrix will be available.

%Other observables  are sensitive to  the effects  of the new operators in \eqref{heff}. In particular, an interesting observable is the ratio
Observables sensitive to the effects of the new operators in \eqref{heff} can be studied. An observable is the ratio
\be
R_{21s}(w)=\frac{{\hat J}_{2s}(w)}{{\hat J}_{1s}(w)}\label{R21s} \,\,.
\ee

%
%%%%%%
%\begin{figure}[!h]
%\begin{center}
%\includegraphics[width = 0.88\textwidth]{angcoeff.png}
%\caption[Angular coefficient functions.]{\small Angular coefficient functions in Eq.~\eqref{angular} for  $\bar B \to D^*(D\pi) \mu^- \bar \nu_\mu$. The shaded regions correspond to the results obtained using the determined  $\epsilon$ couplings, the points are the Belle measurements \cite{Belle:2023xgj}. }\label{angcoeff}
%\end{center}
%\end{figure}
%%%%%%%%%%
\begin{figure}[!h]
\begin{center}
\includegraphics[width = 0.4\textwidth]{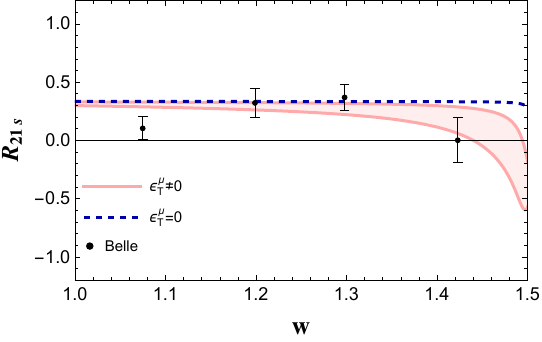}
\includegraphics[width = 0.4\textwidth]{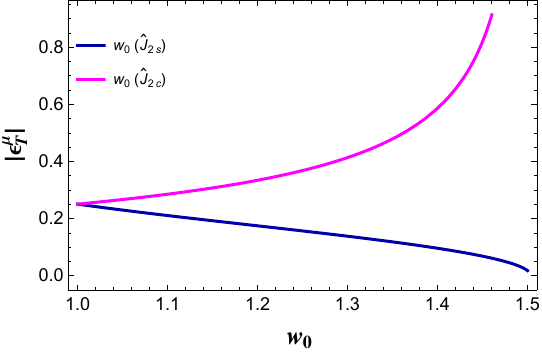}
\caption[$R_{21s}$ observable.]{ \small Left panel: Ratio 
$R_{21s}$ in Eq.~\eqref{R21s} for $\epsilon_T^\mu=0$ (dashed line) and varying $\epsilon_T^\mu$ in the range displayed in Fig.~\ref{Fig:epsilon} (shaded region). Right panel:$|\epsilon_T^\mu |$ as a function of the position of the zero of the ratio ${\hat J}_{2s}$ (blue curve) and  ${\hat J}_{2c}$ (magenta curve) for $\epsilon_V^\mu=\epsilon_R^\mu=0$.
}\label{Fig:2}
\end{center}
\end{figure}
%%%%%%%%%%
%\begin{figure}[!h]
%\begin{center}
%\includegraphics[width = 0.6\textwidth]{plabsepsTnew.pdf}
%\caption[ zero of the ratio ${\hat J}_{2s}$ and ${\hat J}_{2c}$.]{\small  $|\epsilon_T^\mu |$ as a function of the position of the zero of the ratio ${\hat J}_{2s}$ (blue curve) and  ${\hat J}_{2c}$ (magenta curve) for $\epsilon_V^\mu=\epsilon_R^\mu=0$.
%}\label{Fig:absepsT}
%\end{center}
%\end{figure}
%
\noindent
The results in \cite{Colangelo:2024mxe} show that the functions ${\hat J}_{1s,\,2s}$ do not depend on $\epsilon_P$. For $\epsilon_T \neq 0$ their ratio is independent of the form factors and insensitive to $\epsilon_V,\,\epsilon_R$. Therefore, the ratio \eqref{R21s} can signal the tensor operator. This is displayed in Fig.~\ref{Fig:2}, left panel: for non-vanishing $\epsilon_T^{\mu}$ the ratio can have a zero $w_0 \in [1.44,\,1.5]$.

\begin{figure}[!h]
\begin{center}
\includegraphics[width = 0.4\textwidth]{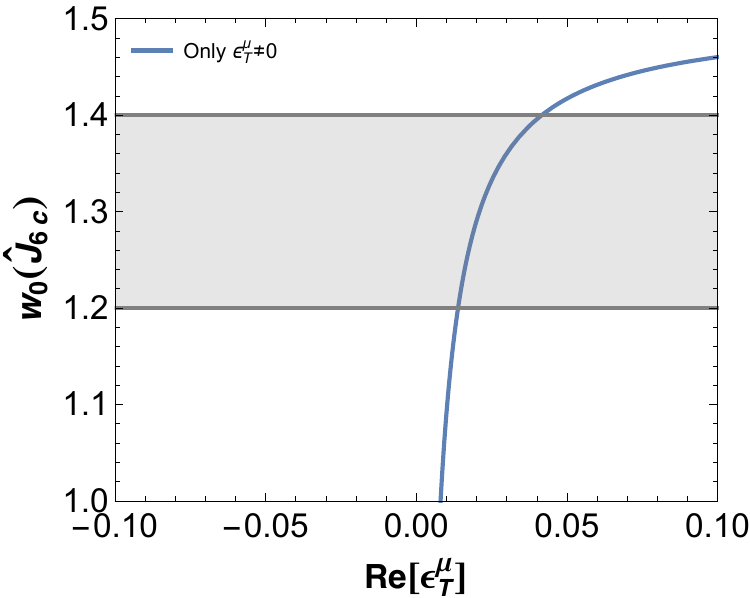}
\caption[ zero of  ${\hat J}_{6c}$.]{\small  Position of the zero of ${\hat J}_{6c}$ vs ${\rm Re}[\epsilon_T^\mu]$, for $\epsilon_V^\mu=\epsilon_R^\mu=\epsilon_P^\mu=0$, obtained from Eq.~(\ref{I6c}).
}\label{zeroI6c}
\end{center}
\end{figure}

The angular coefficient functions provide insights on the possibility that $\epsilon_T$ is the only non-vanishing new coupling. In particular, for $\epsilon_V=\epsilon_R=0$ one finds that both  ${\hat J}_{2s}$  and  ${\hat J}_{2c}$ can have a zero, Fig.~\ref{Fig:2} right panel. If $|\epsilon_T|<0.25$, ${\hat J}_{2s}$ should have a zero while ${\hat J}_{2c}$ should not, and viceversa.  They cannot have a zero simultaneously. 
Although the Belle data  are not precise enough to draw definite conclusions,  they seem to exclude the presence of a zero in ${\hat J}_{2c}$ and are compatible with the presence of a zero of ${\hat J}_{2s}$ in the last bin of $w$. Other angular coefficient functions provide us with further information. If the only non-vanishing NP coupling  is $\epsilon_T$, ${\hat J}_{6c}$ would display a zero  at a value given by the relation
\be
\sqrt{q^2}H_L^{\np}(q^2) {\rm Re}[\epsilon_T]-4m_\ell H_0(q^2)=0\,\,\, . \label{I6c}
\ee
% with $q^2=\displaystyle\frac{m_B^2+m_{D^*}^2-2m_B m_{D^*}w}{2m_B m_{D^*}}$.
The position $w_0$ of the zero of ${\hat J}_{6c}$ would fix ${\rm Re}[\epsilon_T]$. This is shown in Fig.~\ref{zeroI6c}: the continuous curve corresponds to the relation ~\eqref{I6c}, the gray band is  the range of $w$ where the zero of ${\hat J}_{6c}$ should be found according to the Belle results. 

%%%%%%%%%%
%%%%%
%
\section{Summary}
We have constrained the NP coefficients in the generalized low-energy Hamiltonian using a measurement of the full set of angular coefficient functions in $\bar B \to D^* (D \pi)  \mu \bar \nu_\mu$. The possibility that some NP coefficients differ from zero emerges as a preliminary indication based on the data in \cite{Belle:2023xgj}. This observation can be further validated once more complete experimental data will be available.

\vspace*{0.5cm}
\noindent {\bf Acknowledgements.}
I thank P. Colangelo, F. De Fazio and F. Loparco for collaboration. This study has been carried out within the INFN project (Iniziativa Specifica) SPIF.
The research has been partly funded by the European Union – Next Generation EU through the research Grant No. P2022Z4P4B “SOPHYA - Sustainable Optimised PHYsics Algorithms: fundamental physics to build an advanced society" under the program PRIN 2022 PNRR of the Italian Ministero dell’Università e Ricerca (MUR).

\bibliographystyle{JHEP}
\bibliography{my-bib-database}

\providecommand{\href}[2]{#2}\begingroup\raggedright\begin{thebibliography}{10}

\bibitem{DeFazio:2023lmy}
F.~De~Fazio, {\it {Novelties from Flavour Physics}},  11, 2023.
\newblock \href{http://arxiv.org/abs/2311.02987}{{\tt arXiv:2311.02987}}.

\bibitem{Colangelo:2024ped}
P.~Colangelo, F.~De~Fazio, F.~Loparco, and N.~Losacco, {\it {Flavour anomalies,
  correlations, hadronic uncertainties, and all that}},  {\em Nuovo Cim. C}
  {\bf 47} (2024), no.~4 145, [\href{http://arxiv.org/abs/2401.02796}{{\tt
  arXiv:2401.02796}}].

\bibitem{HFLAV:2022pwe}
Y.~Amhis et~al., {\it {Averages of $b$-hadron, $c$-hadron, and $\tau$-lepton
  properties as of 2021}},  {\em Phys. Rev. D} {\bf 107} (2023) 052008,
  [\href{http://arxiv.org/abs/2206.07501}{{\tt arXiv:2206.07501}}].

\bibitem{Colangelo:2016ymy}
P.~Colangelo and F.~De~Fazio, {\it {Tension in the inclusive versus exclusive
  determinations of $|V_{cb}|$: a possible role of new physics}},  {\em Phys.
  Rev.} {\bf D95} (2017) 011701, [\href{http://arxiv.org/abs/1611.07387}{{\tt
  arXiv:1611.07387}}].

\bibitem{Colangelo:2018cnj}
P.~Colangelo and F.~De~Fazio, {\it {Scrutinizing $ \overline{B}\to
  {D}^{\ast}\left(D\pi \right){\ell}^{-}{\overline{\nu}}_{\ell } $ and $
  \overline{B}\to {D}^{\ast}\left(D\gamma
  \right){\ell}^{-}{\overline{\nu}}_{\ell } $ in search of new physics
  footprints}},  {\em JHEP} {\bf 06} (2018) 082,
  [\href{http://arxiv.org/abs/1801.10468}{{\tt arXiv:1801.10468}}].

\bibitem{Bhattacharya:2018kig}
S.~Bhattacharya, S.~Nandi, and S.~Kumar~Patra, {\it {$b \rightarrow c \tau \nu
  _{\tau }$ Decays: a catalogue to compare, constrain, and correlate new
  physics effects}},  {\em Eur. Phys. J. C} {\bf 79} (2019), no.~3 268,
  [\href{http://arxiv.org/abs/1805.08222}{{\tt arXiv:1805.08222}}].

\bibitem{Murgui:2019czp}
C.~Murgui, A.~Pe\~nuelas, M.~Jung, and A.~Pich, {\it {Global fit to $b \to c
  \tau \nu$ transitions}},  {\em JHEP} {\bf 09} (2019) 103,
  [\href{http://arxiv.org/abs/1904.09311}{{\tt arXiv:1904.09311}}].

\bibitem{Becirevic:2019tpx}
D.~Be\v{c}irevi\'c, M.~Fedele, I.~Ni\v{s}and\v{z}i\'c, and A.~Tayduganov, {\it
  {Lepton Flavor Universality tests through angular observables of
  $\overline{B}\to D^{(\ast)}\ell\overline{\nu}$ decay modes}},
  \href{http://arxiv.org/abs/1907.02257}{{\tt arXiv:1907.02257}}.

\bibitem{Bobeth:2021lya}
C.~Bobeth, M.~Bordone, N.~Gubernari, M.~Jung, and D.~van Dyk, {\it
  {Lepton-flavour non-universality of ${\bar{B}}\rightarrow D^*\ell {{\bar{\nu
  }}}$ angular distributions in and beyond the Standard Model}},  {\em Eur.
  Phys. J. C} {\bf 81} (2021), no.~11 984,
  [\href{http://arxiv.org/abs/2104.02094}{{\tt arXiv:2104.02094}}].

\bibitem{Martinelli:2021onb}
G.~Martinelli, S.~Simula, and L.~Vittorio, {\it {$\vert V_{cb} \vert$ and
  $R(D)^{(*)}$) using lattice QCD and unitarity}},  {\em Phys. Rev. D} {\bf
  105} (2022), no.~3 034503, [\href{http://arxiv.org/abs/2105.08674}{{\tt
  arXiv:2105.08674}}].

\bibitem{Belle:2023xgj}
{\bf Belle} Collaboration, M.~T. Prim et~al., {\it {Measurement of Angular
  Coefficients of $\bar{B} \to D^* \ell \bar{\nu}_\ell$: Implications for
  $|V_{cb}|$ and Tests of Lepton Flavor Universality}},
  \href{http://arxiv.org/abs/2310.20286}{{\tt arXiv:2310.20286}}.

\bibitem{Buchmuller:1985jz}
W.~Buchmuller and D.~Wyler, {\it {Effective Lagrangian Analysis of New
  Interactions and Flavor Conservation}},  {\em Nucl. Phys. B} {\bf 268} (1986)
  621--653.

\bibitem{Grzadkowski:2010es}
B.~Grzadkowski, M.~Iskrzynski, M.~Misiak, and J.~Rosiek, {\it {Dimension-Six
  Terms in the Standard Model Lagrangian}},  {\em JHEP} {\bf 1010} (2010) 085,
  [\href{http://arxiv.org/abs/1008.4884}{{\tt arXiv:1008.4884}}].

\bibitem{Colangelo:2024sbf}
P.~Colangelo, F.~De~Fazio, F.~Loparco, and N.~Losacco, {\it {Constraining
  \ensuremath{\nu}SMEFT coefficients: The case of the extra U(1)'}},  {\em
  Phys. Rev. D} {\bf 110} (2024), no.~3 035007,
  [\href{http://arxiv.org/abs/2406.07059}{{\tt arXiv:2406.07059}}].

\bibitem{Colangelo:2024mxe}
P.~Colangelo, F.~De~Fazio, F.~Loparco, and N.~Losacco, {\it {New physics
  couplings from angular coefficient functions of
  B\textasciimacron{}\textrightarrow{}D*(D\ensuremath{\pi})\ensuremath{\ell}\ensuremath{\nu}\textasciimacron{}\ensuremath{\ell}}},
  {\em Phys. Rev. D} {\bf 109} (2024), no.~7 075047,
  [\href{http://arxiv.org/abs/2401.12304}{{\tt arXiv:2401.12304}}].

\bibitem{Navas:PDG}
{\bf Particle Data Group} Collaboration, S.~Navas and Others, {\it {Review of
  Particle Physics}},  {\em Phys. Rev. D} {\bf 110} (2024) 030001.

\end{thebibliography}\endgroup

\end{document}